\documentclass[final,authoryear,5p]{elsarticle}

\usepackage{epsfig}

\usepackage{amssymb}

\usepackage[ps2pdf,%
a4paper=true,%
breaklinks=true,%
colorlinks=true,%
pdfauthor={Hamdi et al.},%
pdftitle={Stark widths for Ar III spectral lines}%
]{hyperref}

\journal{Advances in Space Research}

\begin{document}

\begin{frontmatter}



\title{Stark widths of Ar III spectral lines in the atmospheres of subdwarf B stars}


\author{Rafik Hamdi\corref{cor}}
\address{Groupe de Recherche en Physique Atomique et Astrophysique,
 Facult\'{e} des Sciences de Bizerte, \\ Universit\'{e}
de Carthage, Tunisia.} \cortext[cor]{Corresponding
author}\fnref{footnote1}
\ead{Rafik.Hamdi@istls.rnu.tn}


\author{Nabil Ben Nessib}
\address{Department of Physics and Astronomy, College of Science, King Saud University. PO Box 2455,\\
 Riyadh 11451, Saudi Arabia.}
\ead{nbennessib@ksu.edu.sa}

\author{Sylvie Sahal-Br\'{e}chot}
\address{Laboratoire d'\'{E}tude du Rayonnement et de la
Mati\`{e}re en Astrophysique, Observatoire de Paris, UMR CNRS
8112,\\ UPMC, 5 Place Jules Janssen, 92195 Meudon Cedex, France.}
\ead{sylvie.sahal-brechot@obspm.fr}

\author{Milan S. Dimitrijevi\'{c}
}
\address{Astronomical Observatory, Volgina 7, 11060
Belgrade, Serbia.}
\address{Laboratoire d'\'{E}tude du Rayonnement et de la
Mati\`{e}re en Astrophysique, Observatoire de Paris, UMR CNRS
8112,\\ UPMC, 5 Place Jules Janssen, 92195 Meudon Cedex, France.}
\ead{mdimitrijevic@aob.bg.ac.rs}

\fntext[footnote1]{Deanship of the Foundation Year, Department of
Physics, Umm Al-Qura University, Makkah, Kingdom of Saudi Arabia.}

\begin{abstract}

Using semiclassical perturbation approach in impact approximation,
we have calculated Stark widths for 32 spectral lines of doubly
charged argon (Ar III). Oscillator strengths are calculated using
Hartree-Fock method with relativistic correction (HFR) and an
atomic model including 17 configurations. Energy levels are taken
from NIST database. For perturbing levels for which the
corresponding energy does not exist in NIST database, the
calculated energies are used. Our widths are compared with the
experimental results. The results presented here are of interest
for modelling and investigation of stellar atmospheres since argon
in different ionization stages is observed in many astrophysical
objects. Finally, the importance of Stark broadening mechanism is
studied in the atmospheric conditions of sdB stars. Electron
impact Stark widths are compared to thermal Doppler widths as a
function of temperature and optical depth of atmospheric layers.

\end{abstract}


\end{frontmatter}

\parindent=0.5 cm

\section{Introduction}
Stark broadening parameters (width and shift) are of interest for
the study of astrophysical and laboratory plasma. Stark Broadening
parameters can be used in the determination of temperature and
density of laboratory plasma. For example, in \citet{Zhou09},
electron temperature and density are determined simultaneously in
a cold argon arc-plasma jet by using Stark broadening of two
different emission lines.

Stark broadening is important for modelling and investigation of
stellar atmospheres of A and B stars \citep{Popovic01,Simic05}.
\citet{Dimitri07} studied the Stark broadening on the line shapes
of Cr II spectral lines observed in stellar atmospheres of middle
part of the main sequence. They found that Stark broadening
mechanism is very important and should be taken into account,
especially in the study of Cr abundance stratification.

Besides main sequence stars, Stark broadening mechanism is
important for white dwarfs. \citet{Hamdi08} considered the
broadening on Si VI lines in DO white dwarf spectra. They found
that Stark broadening is dominant in broad regions of the
considered DO atmospheres. For much cooler DB white dwarfs, Stark
broadening is usually the dominant broadening mechanism
\citep{dimitri11,Simic09}.

Doubly charged Argon (Ar III) spectral lines are observed in many
astrophysical plasmas. In \citet{Rodrig99}, Ar III lines are used
in the determination of abundance in galactic H II regions.
\citet{Blanchette08}, used Ar III $\lambda$ 1002.097 $\rm\AA$ line
in the determination of abundance in hydrogen-riche subdwarf B
stars. \citet{Otoole06} obtained high-resolution ultraviolet
spectra of five sdB stars using \textit{Space Telescope Imaging
Spectrograph} onboard the \textit{Hubble Space Telescope}.
Abundance of Ar III ion was determined in the studied sdB stars.

Stark widths measurement of Ar III spectral lines are reported in
eight works:
\citet{Platisa75,Baker79,Konjevic87,Puric88,Konjevic90,Djenize96,Bukvic08,Djurovic11}.
 In \citet{Baker79}, Stark broadening parameters were
determined in 870-890 $\rm\AA$ wavelength interval. In all other
works, Stark broadening parameters were determined in 2140-3960
$\rm\AA$ wavelength interval.

In this work, we have calculated Stark widths for 32 Ar III
spectral lines. We have used semiclassical perturbation approach
in impact approximation \citep{ssb69a,ssb69b}. Energy levels are
taken from NIST database \citep{Ralchenko11} and oscillator
strengths are calculated using Cowan code \citep{Cowan81}. Our
Stark widths are compared with the  experimental results of
\citep{Djurovic11,Bukvic08,Djenize96,Konjevic90,Konjevic87,Platisa75}.
Finally, the importance of collisions with electrons in
atmospheric conditions of subdwarf B (sdB) stars is studied.
Electron impact Stark widths are compared with thermal Doppler
width as a function of optical depth and as a function of the
temperature of atmospheric layers.

\section{The impact semiclassical perturbation method}
The impact semiclassical perturbation formalism is described in
\citet{ssb69a,ssb69b}. The innovations to this formalism are given
in \citet{ssb74,ssb91,Fleurier77,DS96}. For example in
\citet{ssb74} the expression of the quadrupole term for complex
atoms was given. The profile $F(\omega )$ is Lorentzian for
isolated lines:
\begin{eqnarray}
F(\omega )=\frac{w/\pi }{(\omega -\omega _{if}-d)^{2}+w^{2}}
\end{eqnarray}
where
\begin{eqnarray*}
\omega _{if}=\frac{E_{i}-E_{f}}{\hbar }
\end{eqnarray*}
$i$ and $f$ denote the initial and final states and $E_{i}$ and
$E_{f}$ their corresponding energies.

The total width at half maximum ($W=2w$) and shift ($d$) (in
angular frequency units) of an electron-impact broadened spectral
line can be expressed as:
\begin{eqnarray*}
W &=&N\int vf(v)dv\left( \sum\limits_{i^{\prime }\neq i}\sigma
_{ii^{\prime }}(v)+\sum\limits_{f^{\prime }\neq f}\sigma
_{ff^{\prime }}(v)+\sigma
_{el}\right) \\
\end{eqnarray*}
\begin{eqnarray}
d &=&N\int vf(v)dv\int\nolimits_{R_{3}}^{R_{D}}2\pi \rho d\rho
\sin (2\varphi _{p})
\end{eqnarray}
where $N\ $is the electron density, $f(\upsilon )\ $the Maxwellian
velocity distribution function for electrons, $\rho \ $denotes the
impact parameter of the incoming electron, $i^{\prime }$ (resp.
$f\ ^{\prime }$) denotes the perturbing levels of the initial
state $i$ (resp. final state $f$). The inelastic cross section
$\sigma _{ii^{\prime }}(\upsilon )$ (resp. $\sigma _{ff^{\prime
}}(\upsilon )$) can be expressed by an integral over the impact
parameter $\rho \ $ of the transition probability $P_{ii^{\prime
}}(\rho ,\upsilon )\ $(resp. $P_{ff^{\prime }}(\rho ,\upsilon )\
$) as
\begin{equation}
\sum_{i^{\prime }\neq i}\sigma _{ii^{\prime }}(\upsilon
)=\frac{1}{2}\pi R_{1}^{2}+\int_{R_{1}}^{R_{D}}2\pi \rho d\rho
\sum_{i^{\prime }\neq i}P_{ii^{\prime }}(\rho ,\upsilon ).
\end{equation}
and the elastic contribution is given by
\begin{eqnarray*}
\sigma _{el}=2\pi R_{2}^{2}+\int_{R_{2}}^{R_{D}}2\pi \rho d\rho
\sin ^{2}\delta+\sigma _{r},
\end{eqnarray*}
\begin{equation}
\delta =(\varphi _{p}^{2}+\varphi _{q}^{2})^{\frac{1}{2}}.
\end{equation}
The phase shifts $\varphi _{p}\ $and$\ \varphi _{q}\ \ $due
respectively to the polarization potential ($r^{-4}$) and to
the quadrupolar potential ($%
r^{-3}$), are given in Section 3 of Chapter 2 in \citet{ssb69a}
and $R_{D}$ is the Debye radius. All the cut-offs $R_{1}$,$\
R_{2}$ and$\ R_{3}$ are described in Section 1 of Chapter 3 in
\citet{ssb69a}. $\sigma _{r}$ is the contribution of the Feshbach
resonances \citep{Fleurier77}.

The formulae for the ion-impact widths and shifts are analogous to
Eqs. (2)-(4), without the Feshbach resonances contribution to the
width. For electrons, hyperbolic paths due to the attractive
Coulomb force are used, while for perturbing ions the hyperbolic
paths are different since the force is repulsive.

Semiclassical perturbation calculations need a relatively large
set of oscillator strengths. In this work, oscillator strengths
are calculated with the Hartree-Fock relativistic approach using
Cowan code \citep{Cowan81} and an atomic model including 17
configurations: 3s$^{2}$3p$^{4}$, 3s$^{2}$3p$^{3}$$nl$ ($nl$=4p,
4f, 5p, 5f, 6p, 6f) (even parity) and 3s3p$^{5}$,
3s$^{2}$3p$^{3}$$n'l'$ ($n'l'$=3d, 4s, 4d, 5s, 5d, 5g, 6s, 6d, 6g)
(odd parity).
\section{Semiclassical perturbation Stark widths and comparison with experiments}

\begin{figure}

\begin{center}
\includegraphics*[width=10cm]{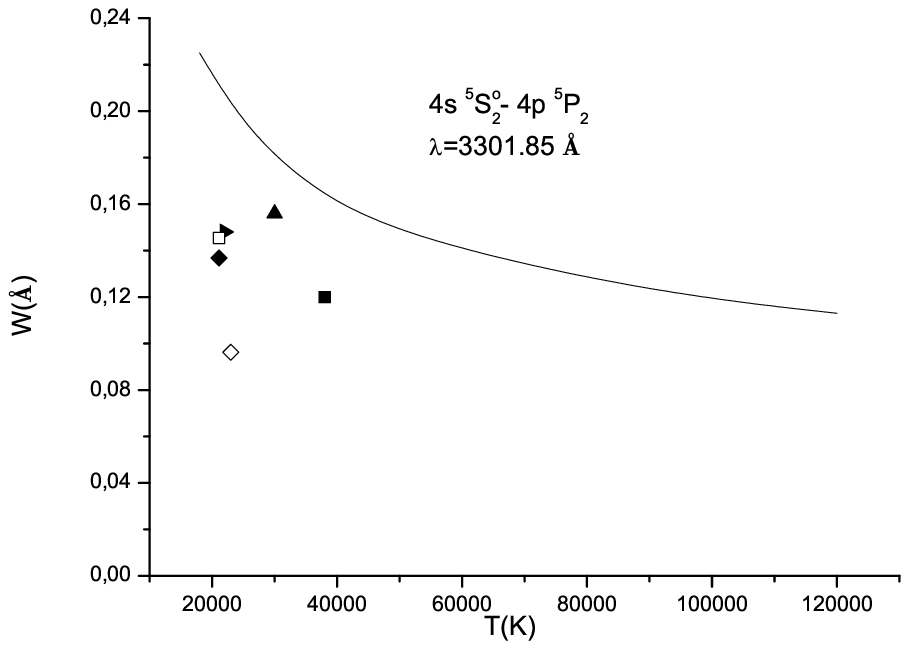}
\end{center}
\caption{Electron impact Stark width (FWHM) for the $4s$
$^{5}$S$^{\rm o}_{2}$ - $4p$ $^{5}$P$_{2}$ ($\lambda$= 3301.85
\AA) line as a function of electron temperature at an electron
density of $10^{17}$ cm$^{-3}$. Solid line: Our Stark widths
obtained using semiclassical perturbation approach
\citep{ssb69a,ssb69b}.( $\blacktriangleright$): experimental Stark
width of \citet{Djurovic11}. ($\blacktriangle$): experimental
Stark width of \citet{Bukvic08}. ($\blacksquare$): experimental
Stark width of \citet{Djenize96}. ($\square$ and $\vartriangle$):
experimental Stark widths of \cite{Konjevic87} given for
temperatures 21100 K and 26000 K respectively. ($\blacklozenge$
and $\lozenge$): experimental Stark widths of \cite{Platisa75}
given for temperatures 21100 K and 23080 K respectively.}
\label{figure1}
\end{figure}

\begin{figure}
\begin{center}
\includegraphics*[width=10cm]{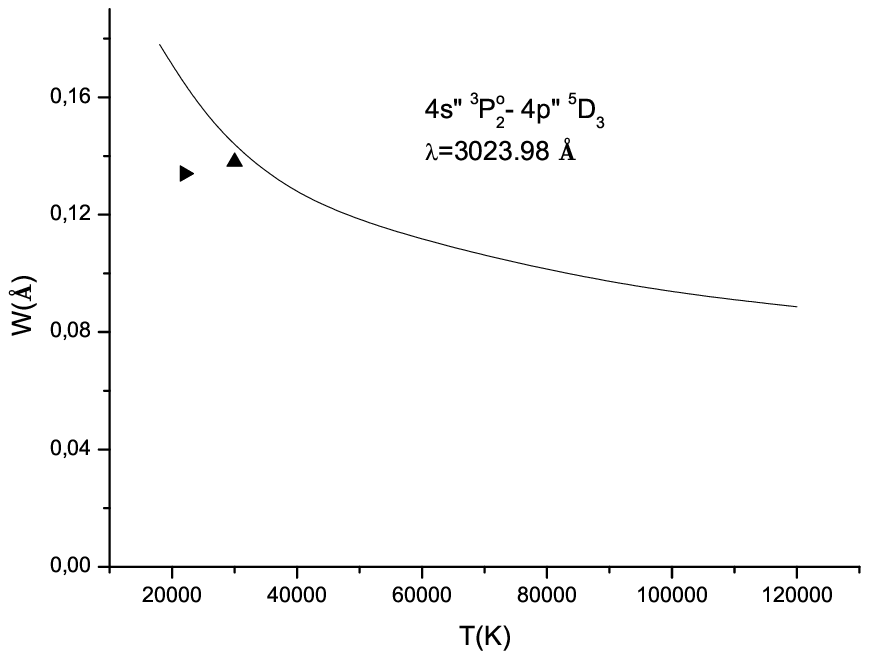}
\end{center}
\caption{Same as in Fig. \ref{figure1} but for the $4s''$
$^{3}$P$^{\rm o}_{2}$ - $4p''$ $^{5}$D$_{3}$ ($\lambda$= 3023.98
\AA) transition.} \label{figure2}
\end{figure}


\citet{Djurovic11} reported Stark width measurements of 19 Ar III
spectral lines. The plasma source was a low-pressure-pulsed arc.
Electron densities of (3.5 - 9.0) $\times$ 10 $^{16}$ cm$^{-3}$
were determined by two-wavelength interferometry method and
electron temperature of 16 000 - 24 000 K was measured with the
help of the Boltzmann plot technique. All spectral lines for which
\citet{Djurovic11} measured Stark parameters belong to the UV
region of 2630 - 3960 $\rm\AA$. Authors report that the errors of
the measured widths vary from 15\% to 50\%. In \citet{Bukvic08},
Stark widths measurements of 12 Ar III spectral lines are
reported. The plasma source was a mixture of Ar (72 \%) and He (28
\%) plasma created in the linear, low pressure, arc discharge. The
electron temperature belonging to the interval (26 000 - 30 000 K)
was estimated using Boltzmann plot technique. The electron density
belonging to the interval (0.16 - 1.68) $\times$ 10 $^{17}$
cm$^{-3}$ was determined using single laser interferometry
technique at the wavelength 6328 $\rm\AA$ of the He-Ne laser.
According to \citet{Bukvic08} the Stark widths were measured with
12 \% error. \citet{Djenize96} reported Stark width measurements
of 13 Ar III spectral lines at an electron density of 3.5 $\times$
10$^{17}$ cm$^{-3}$ and electron temperature of 38 000 K. The
plasma source was also an argon-helium mixture. Electron density
was also measured using laser interferometry at 6328 $\rm\AA$.
Electron temperature was measured using Boltzmann plot and ratios
of Ar IV to Ar III lines and Ar III to Ar II lines. The
uncertainty of the results given by \citet{Djenize96} vary from
$\pm$ 14 \% to $\pm$ 18 \%. \citet{Konjevic90} gives electron
impact widths for six Ar III spectral lines. A low-pressure pulsed
arc was used as plasma source. Electron-density was measured using
Stark width of He II Pashen-$\alpha$ 4684.7 $\rm\AA$ line.
Boltzmann plot of O III 3754.21, 3707.24, 3707.75 and 3455.12
$\rm\AA$ lines was used for the determination of electron
temperature. In \citet{Konjevic87}, electron impact widths for 11
Ar III spectral lines were reported. Electron density was
determined using a He-Ne laser quadrature interferometer operating
at 6328 $\rm\AA$. The ratio of 4366.90 and 4369.28 $\rm\AA$ O II
impurity lines determined electron temperature. \citet{Konjevic87}
estimated an error of $\pm$ 15 \% for the widths.
\citet{Platisa75} reported Stark widths of five Ar III spectral
lines. Electron density was determined using the same method as in
\citet{Konjevic87} and electron temperature were determined from
the Botltzmann plot of relative intensities of eight Ar II lines.
\citet{Platisa75} estimated that the error of the measured widths
is $\pm$ 30 \%.

In Table \ref{table1}, we present our electron impact (W$_{e}$)
and ion impact (W$_{i}$) Stark widths (FWHM) with the
experimentally determined Stark width (W$_{m}$) taken from
\citet{Djurovic11,Bukvic08,Djenize96,Konjevic90,Konjevic87,Platisa75}.
Our Stark widths are calculated using semiclassical perturbation
approach in impact approximation \citep{ssb69a,ssb69b}. Energy
levels needed for this calculation are taken from NIST database
\citep{Ralchenko11} and oscillator strengths are calculated using
Cowan code \citep{Cowan81} and the atomic model previously
described. For perturbing levels for which the corresponding
energy do not exist in NIST database \citep{Ralchenko11}, we have
used the energies that we have calculated using Cowan code
\citep{Cowan81}. In NIST database we have found 125 energy levels
and 497 classified electric dipole transitions. Among these
transitions, only 68 are given with the corresponding oscillator
strengths. This number of oscillator strengths is not sufficient
to perform a semiclassical perturbation calculation of 32 Stark
widths which need a large number of oscillator strengths. So the
use of Cowan code for the calculation of oscillator strengths is
very interesting. In \citet{Hamdi13} and \citet{Hamdi11}, we have
used this method for Pb IV and we have determined Stark broadening
parameters for 114 spectral lines. In our calculation, only
electric dipole (E1) transitions are taken into account. All
wavelengths given in Table \ref{table1} are taken from NIST
database \citep{Ralchenko11}. For each value given in Table
\ref{table1}, the collision volume ($V$) multiplied by perturber
density ($N$) is much less than one and the impact approximation
is valid. In some cases, 0.1$<$\emph{N}\emph{V}$\leq$0.5 and the
impact approximation reaches its limit of validity, these values
are preceded by an asterisk. The greatest value of
$\emph{N}\times\emph{V}$, equal to 0.20 has been found for 4p$'$
$^{3}$D$_{3}^{\rm{o}}$ - 4d$'$ $^{3}$P$_{2}$ transition for
collisions with ions. For each transition given in Table
\ref{table1}, our semiclassical perturbation Stark width is
W$_{sc}$ = W$_{e}$+W$_{i}$. Taking into account the plasma
composition for each experiment, we have taken as ionic perturber
singly ionized helium when we compared with \citet{Bukvic08} and
\citet{Djenize96}, singly ionized argon when we compared with
\citet{Djurovic11} and singly ionized nitrogen when we compared
with \cite{Platisa75}. In \cite{Konjevic90,Konjevic87}, electron
impact widths were reported, for this reason broadening by
collision with ions is not given when we compare with those two
authors.

\begin{figure}
\begin{center}
\includegraphics*[width=10cm]{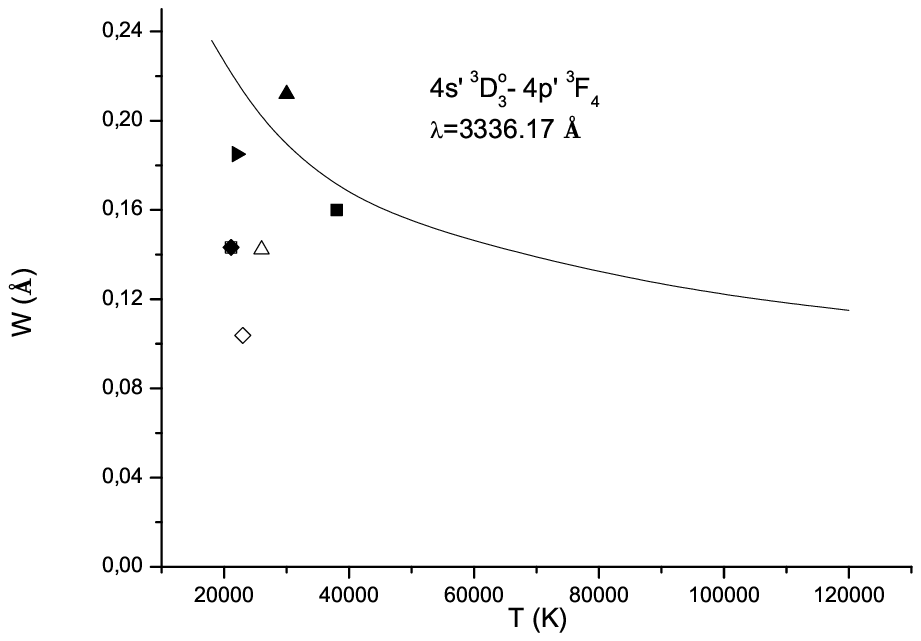}
\end{center}
\caption{Same as in Fig. \ref{figure1} but for the $4s'$
$^{3}$D$^{\rm o}_{3}$ - $4p'$ $^{3}$F$_{4}$ ($\lambda$= 3336.17
\AA) transition.} \label{figure3}
\end{figure}

\begin{figure}
\begin{center}
\includegraphics*[width=10cm]{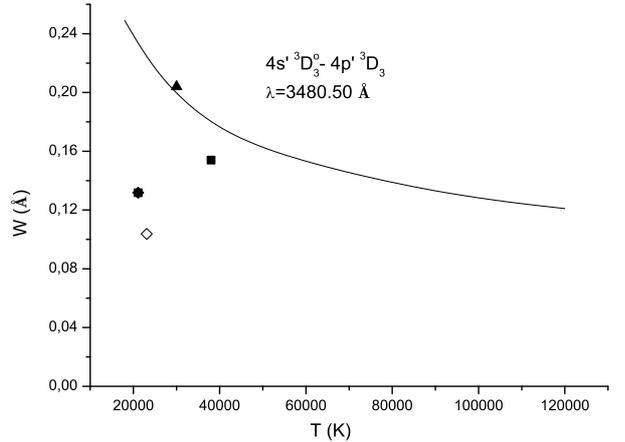}
\end{center}
\caption{Same as in Fig. \ref{figure1} but for the $4s'$
$^{3}$D$^{\rm o}_{3}$ - $4p'$ $^{3}$D$_{3}$ ($\lambda$= 3480.50
\AA) transition.} \label{figure4}
\end{figure}

The agreement of our Stark widths with \citet{Djurovic11} values
is 35\% in average. All our values are greater than
\citet{Djurovic11} ones. The lower value of ratio
$\frac{W_{m}}{W_{sc}}$ equal to 0.93 is found for the 3d$'$
$^{3}$D$_{1}^{\rm{o}}$ - 4p$'$ $^{3}$D$_{1}$ transition. The
largest differences between our values and experimental Stark
widths of \citet{Djurovic11} are found for 4s$'$ - 4p$'$
transitions as for example for the 4s$'$ $^{3}$D$_{2}^{\rm{o}}$ -
4p$'$ $^{3}$P$_{2}$ transition for which the ratio
$\frac{W_{m}}{W_{sc}}$ is equal to 0.67. We notice that this
transition is classified by \citet{Djurovic11} as C (errors up to
50\%). So, taken into account the accuracy of the experimental
results, the agreement with Stark widths of \citet{Djurovic11} is
acceptable for all transitions. Very good agreement is found with
\citet{Bukvic08} (8.6 \% on average). The greatest difference
(31\%) between our Stark widths and \citet{Bukvic08} ones is found
for 4s$''$ $^{3}$P$_{2}^{\rm{o}}$ - 4p$''$ $^{3}$D$_{2}$
transition. For the 3d$''$ $^{3}$P$_{1}^{\rm{o}}$ - 4p$''$
$^{3}$P$_{1}$ transition the ratio $\frac{W_{m}}{W_{sc}}$ is equal
to 1.  On average, our Stark widths agree with \citet{Djenize96}
ones within 26\%. Besides multiplets arising from 4s, 4p, 4s$'$,
4p$'$, 3d$''$, 4p$''$ parent energy levels, results for higher
multiplets arising from 5s, 4d$'$, 5s$'$ are also given. An
agreement better than 30\% is found for these multiplets except
for 4p$'$ $^{3}$D$_{3}$ - 4d$'$ $^{3}$P$_{2}^{\rm{o}}$ transition,
the difference between our calculated Stark width and the measured
one is 46\%. The agreement between our electron impact Stark
widths and the values of \citet{Konjevic90} is within 35\%. In
\citet{Konjevic90}, Stark widths are given for two experimental
conditions: T = 80 000 K, Ne = 5.8 $\times$ 10$^{17}$ cm$^{-3}$
and T = 110 000 K, Ne = 1 $\times$ 10$^{18}$ cm$^{-3}$. Our
results agree better with the experimental Stark widths given for
T = 80 000 K and Ne = 5.8 $\times$ 10 $^{17}$ cm$^{-3}$. All our
Stark widths are greater than \citet{Konjevic90} ones. Our
electron impact Stark widths agree with \citet{Konjevic87} ones
within 40 \%. All our results are greater than \citet{Konjevic87}
ones. The greatest difference (76\%) between our Stark widths and
\citet{Konjevic87} values is found for the $4s'$ $^{3}$D$^{\rm
o}_{3}$ - $4p'$ $^{3}$D$_{3}$ ($\lambda$= 3480.50 \AA) transition.
In \citet{Platisa75}, measured Stark widths are given for two
experimental conditions: T = 21 100 K, Ne = 0.44 $\times$ 10
$^{17}$ cm$^{-3}$ and T = 23 080 K, Ne = 0.80 $\times$ 10 $^{17}$
cm$^{-3}$. The agreement of our widths with the results of
\citet{Platisa75} is not good specially for T = 23 080 K and Ne =
0.80 $\times$ 10 $^{17}$ cm$^{-3}$. For some lines we found a
factor greater than two between our Stark widths and
\citet{Platisa75} ones.

In Fig. \ref{figure1}, we present our electron impact Stark width
as a function of electron temperature for the interval (20 000 -
120 000 K) along with experimental values of \citet{Djurovic11};
\citet{Bukvic08}; \citet{Djenize96}; \citet{Konjevic90};
\citet{Konjevic87} and \citet{Platisa75} for the transition $4s$
$^{5}$S$^{\rm o}_{2}$ - $4p$ $^{5}$P$_{2}$ ($\lambda$= 3301.85
\AA). All experimental values are normalized to an electron
density of $10^{17}$ cm$^{-3}$. This figure shows that our results
over estimate all the experimental values. Fig. \ref{figure2},
Fig. \ref{figure3} and Fig. \ref{figure4}, are the same as Fig.
\ref{figure1} but for the transitions: $4s''$ $^{3}$P$^{\rm
o}_{2}$ - $4p''$ $^{5}$D$_{3}$ ($\lambda$= 3023.98 \AA),  $4s'$
$^{3}$D$^{\rm o}_{3}$ - $4p'$ $^{3}$F$_{4}$ ($\lambda$= 3336.17
\AA) and $4s'$ $^{3}$D$^{\rm o}_{3}$ - $4p'$ $^{3}$D$_{3}$
($\lambda$= 3480.50 \AA) respectively. Fig. \ref{figure2} shows
also that our theoretical Stark widths overestimate the
experimental values. We can see also that our width at T = 30 000
K is very close to \citet{Bukvic08} value. In Fig. \ref{figure3},
we can see that our electron impact Stark width underestimate the
experimental value of \citet{Bukvic08} and overestimate all others
results. The result of \citet{Djenize96} is the closest to our
value. Fig. \ref{figure4} shows that our widths overestimate the
experimental values of \citet{Djenize96,Konjevic87,Platisa75} and
that our width at T = 30 000 K is very close to \citet{Bukvic08}
one. Fig. \ref{figure1}, Fig. \ref{figure3} and Fig. \ref{figure4}
show a large difference between our widths and \citet{Platisa75}
ones at T = 23080 K.

Our results as a function of temperature and electron density will
be published elsewhere and will be inserted in the Stark-B
database \citep{ssb12}, which is a part of Virtual Atomic and
Molecular Data Center \citep{Dubernet10}. Besides the study and
investigation of stellar atmospheres, this database is also
devoted to the study of laboratory an fusion plasma. For example,
Ar III ion spectral lines are observed by \citet{Graf11} in the
spectrum of deuterium plasma in the Alcator C-Mod Tokamak.
\begin{table*}
\begin{center}
\caption{Our electron impact Stark widths (FWHM) (W$_{e}$) and ion
impact Stark widths (W$_{i}$) calculated using SCP approach in
impact approximation \citep{ssb69a,ssb69b} compared with
experimental values of \citet{Djurovic11,Bukvic08,Djenize96}
(W$_{m}$). Transitions, wavelengths, electron temperature (T) and
electron density (N$_{e}$) are also given. Ref.: a:
\citet{Djurovic11}; b: \citet{Bukvic08}; c: \citet{Djenize96}; d:
\cite{Konjevic90}; e: \cite{Konjevic87}; f: \cite{Platisa75}.}
\begin{tabular}{cccrccccc}
\hline Transition & Term & $\lambda $ (\AA) & T (K) & N$_{e}$
(10$^{17}$ cm$^{-3} $) & W$_{m}$ (pm) & W$_{e}$ (pm)& W$_{i}$
(pm)& Ref.
\\ \hline
4s - 4p&$^{5}$S$_{2}^{\rm{o}}$ - $^{5}$P$_{3}$  &3285.84  &22000  &1.00  &14.9  &20.4  &1.12  &a  \\
&  &  &38000  &3.50  &46.0  &56.3  &4.26  &c  \\
&  &  &21100  &0.44  &6.4  &9.13  & &e  \\
&  &  &26000  &0.85  &12.2  &16.1  & &e  \\
&  &  &21100  &0.44  &6.4  &9.13  &0.49 &f  \\
&  &  &23080  &0.80  &8.2  &15.9  &0.92 &f  \\
&$^{5}$S$_{2}^{\rm{o}}$ - $^{5}$P$_{2}$  &3301.85 &22000  &1.00  &14.8  &20.5  &1.22  &a\\
&  &  &30000  &1.47  &23.0  &26.3  &1.66  &b\\
&  &  &38000  &3.50  &42.0  &56.6  &4.30  &c\\
&  &  &21100  &0.44  &6.4  &9.19  &  &e\\
&  &  &21100  &0.44  &6.1  &9.19  &0.49  &f\\
&  &  &23080  &0.80  &7.7  &16.1  &0.93  &f\\
&$^{5}$S$_{2}^{\rm{o}}$ - $^{5}$P$_{1}$  &3311.24 &22000  &1.00
&16.1 &20.6  &1.23  &a\\
&  &  &30000  &1.47  &25.8  &26.4  &1.67  &b\\
&  &  &26000  &0.85  &12.2  &16.2  &  &e\\
4s$'$ - 4p$'$&$^{3}$D$_{1}^{\rm{o}}$ - $^{3}$P$_{1}$  &2853.30  &22000  &1.00  &11.9  &15.4  &1.63  &a  \\
&$^{3}$D$_{2}^{\rm{o}}$ - $^{3}$P$_{1}$  &2855.31  &22000  &1.00  &12.3  &15.4  &1.63  &a  \\
& &  &80000  &5.8  &44.0  &53.9  &  &d  \\
& &  &110000  &10  &62.0  &84.9  &  &d  \\
&$^{3}$D$_{2}^{\rm{o}}$ - $^{3}$P$_{2}$  &2878.76  &22000  &1.00  &11.6  &15.5  &1.64  &a  \\
&$^{3}$D$_{3}^{\rm{o}}$ - $^{3}$P$_{2}$  &2884.21  &22000  &1.00  &11.2  &15.6  &1.65  &a  \\
& &  &80000  &5.8  &45.0  &54.4  &  &d  \\
& &  &110000  &10  &66.0  &85.7  &  &d  \\
4s$'$ - 4p$'$&$^{3}$D$_{2}^{\rm{o}}$ - $^{3}$D$_{3}$  &3472.56  &30000  &1.47  &32.0  &28.7  &2.94  &b  \\
&$^{3}$D$_{3}^{\rm{o}}$ - $^{3}$D$_{3}$  &3480.50  &30000  &1.47  &30.0  &28.9  &2.95  &b  \\
  & &  &38000  &3.50  &54.0  &62.1  &7.32  &c  \\
  & &  &21100  &0.44 &5.8  &10.2  &  &e  \\
  & &  &21100  &0.44 &5.8  &10.2  &0.88  &f  \\
  & &  &23080  &0.80 &8.1  &17.7  &1.63  &f  \\
&$^{3}$D$_{2}^{\rm{o}}$ - $^{3}$D$_{2}$  &3503.58  &38000  &3.50  &48.0  &60.6  &7.38  &c  \\
& &  &80000  &5.80  &53.0  &77.2  &  &d  \\
& &  &110000  &10.0  &78.0  &121  &  &d  \\
& &  &27500  &0.84  &12.4  &16.7  &  &e  \\
4s$'$ - 4p$'$&$^{3}$D$_{3}^{\rm{o}}$ - $^{3}$F$_{4}$  &3336.17  &22000  &1.00  &18.5  &21.5  &1.98  &a  \\
& &  &30000  &1.47  &31.2  &27.5  &2.79  &b  \\
& &  &38000  &3.50  &56.0  &59.0  &6.91  &c  \\
& &  &21100  &0.44  &6.3  &9.64  &  &e  \\
& &  &26000  &0.85  &12.1  &16.9  &  &e  \\
& &  &21100  &0.44  &6.3  &9.64  & 0.83 &f  \\
& &  &23080  &0.80  &8.3  &16.8  & 1.54 &f  \\
&$^{3}$D$_{2}^{\rm{o}}$ - $^{3}$F$_{3}$  &3344.75  &22000  &1.00  &16.6  &21.4  &1.98  &a  \\
&  &  &30000  &1.47  &32.5  &27.4  &2.80  &b  \\
&  &  &38000  &3.50  &54.0  &58.8  &6.92  &c  \\
&  &  &26000  &0.85  &12.1  &16.9  &  &e  \\
&$^{3}$D$_{3}^{\rm{o}}$ - $^{3}$F$_{3}$  &3352.11  &30000  &1.47  &30.9  &27.5  &2.81  &b  \\
&$^{3}$D$_{1}^{\rm{o}}$ - $^{3}$F$_{2}$  &3358.53  &22000  &1.00  &16.0 &21.3  &1.99  &a  \\
&  & &30000  &1.47  & 31.1&27.4  &2.81  &b  \\
&  & &38000  &3.50  & 46.0&58.9  &6.96  &c  \\
&$^{3}$D$_{2}^{\rm{o}}$ - $^{3}$F$_{2}$  &3361.30  &30000  &1.47  &31.8  &27.3  &3.21  &b  \\
4s$''$ - 4p$''$&$^{3}$P$_{2}^{\rm{o}}$ - $^{3}$P$_{2}$  &2762.16  &22000  &1.00  &12.1  &13.9  &*2.07  &a  \\
&$^{3}$P$_{2}^{\rm{o}}$ - $^{3}$P$_{1}$  &2783.60  &22000  &1.00
&13.3  &14.1  &*2.09  &a \\\hline
\end{tabular}
\end{center}
\end{table*}
\newpage
\addtocounter{table}{-1}
\begin{table*}
\caption{$Continued.$}
\label{tab:3}       
\begin{center}
\begin{tabular}{cccrccccc}
\hline Transition & Term & $\lambda $ (\AA) & T (K) & N$_{e}$
(10$^{17}$ cm$^{-3} $) & W$_{m}$ (pm) & W$_{e}$ (pm)& W$_{i}$
(pm)& Ref.
\\ \hline

&$^{3}$P$_{1}^{\rm{o}}$ - $^{3}$P$_{2}$  &2785.20  &22000  &1.00  &12.3  &14.1  &*2.11  &a  \\
&$^{3}$P$_{1}^{\rm{o}}$ - $^{3}$P$_{0}$  &2824.64  &22000  &1.00  &11.7  &14.8  &*2.14  &a  \\
4s$''$ - 4p$''$&$^{3}$P$_{2}^{\rm{o}}$ - $^{3}$D$_{3}$  &3023.98  &22000  &1.00  &13.4  &16.3  &2.28  &a  \\
& &  &30000  &1.47  &20.4  &20.9  &3.22  &b  \\
&$^{3}$P$_{2}^{\rm{o}}$ - $^{3}$D$_{2}$  &3027.08  &30000  &1.47  &18.4  &20.9  &3.22  &b  \\
&$^{3}$P$_{1}^{\rm{o}}$ - $^{3}$D$_{2}$  &3054.77  &22000  &1.00 &15.4  &16.6  &2.32  &a\\
& &  &27500  &0.84 &11.4  &12.7  &  &e\\
3d$'$ - 4p$'$&$^{3}$D$_{1}^{\rm{o}}$ - $^{3}$D$_{1}$  &2631.86  &22000  &1.00  &9.8  &9.61  &1.22  &a  \\
&$^{3}$D$_{2}^{\rm{o}}$ - $^{3}$D$_{2}$  &2678.35  &22000  &1.00  &10.3  &9.70  &1.33  &a  \\
3d$''$ - 4p$''$&$^{3}$P$_{2}^{\rm{o}}$ - $^{3}$S$_{1}$  &3960.49  &22000  &1.00  &21.2  &26.1  &*4.25  &a  \\
3d$''$ - 4p$''$&$^{3}$P$_{2}^{\rm{o}}$ - $^{3}$P$_{2}$  &3391.84  &38000  &3.50  &52.0  &56.6  &12.0  &c  \\
&  & &21100  &0.44  &5.8  &9.04  &1.45  &f  \\
&$^{3}$P$_{1}^{\rm{o}}$ - $^{3}$P$_{1}$  &3471.29  &30000  &1.47  &31.6  &26.6  &5.03  &b  \\
4p - 5s&$^{5}$P$_{1}$ - $^{5}$S$_{2}^{\rm{o}}$  &2166.18  &38000  &3.50  &53.0  &55.1  &4.73  &c  \\
&$^{5}$P$_{2}$ - $^{5}$S$_{2}^{\rm{o}}$  &2170.22  &38000  &3.50  &61.0  &55.3  &4.75  &c  \\
&$^{5}$P$_{3}$ - $^{5}$S$_{2}^{\rm{o}}$  &2177.19  &38000  &3.50  &56.0  &55.8  &4.78  &c  \\
4p$'$ - 4d$'$&$^{3}$D$_{3}$ - $^{3}$P$_{2}^{\rm{o}}$  &2168.28  &38000  &3.50  &69.0  &38.4  &*8.94  &c  \\
4p$''$ - 5s$'$&$^{3}$D$_{3}$ - $^{3}$D$_{3}^{\rm{o}}$  &2133.86
&38000  &3.50 &51.0  &40.4  &3.96  &c  \\ \hline
\end{tabular}
  \label{table1}
\end{center}
\end{table*}

\section{Stark broadening effect in sdB stars atmospheres}

\begin{figure}
\begin{center}
\includegraphics*[width=10cm]{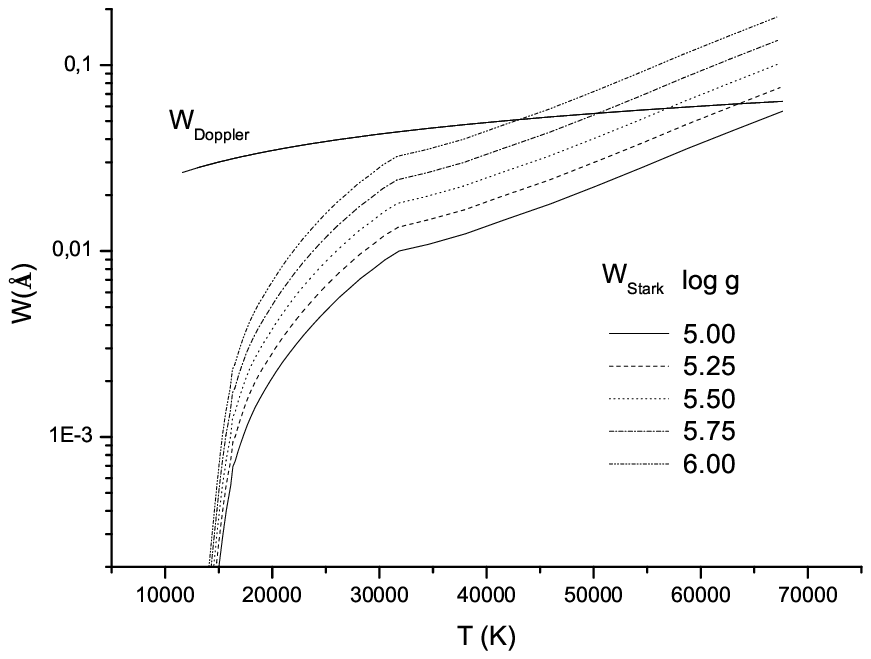}
\end{center}
\caption{Stark and Doppler widths for Ar III 4p $^{5}$P$_{2}$ - 5s
$^{5}$S$_{2}^{\rm{o}}$ ($\lambda$ = 2170.22 \AA) spectral line as
a function of atmospheric layer temperature. Stark widths are
shown for five values of model gravity log \emph{g} = 5-6,
T$_{eff}$ = 22 000 K.} \label{sdb1}
\end{figure}

\begin{figure}
\begin{center}
\includegraphics*[width=10cm]{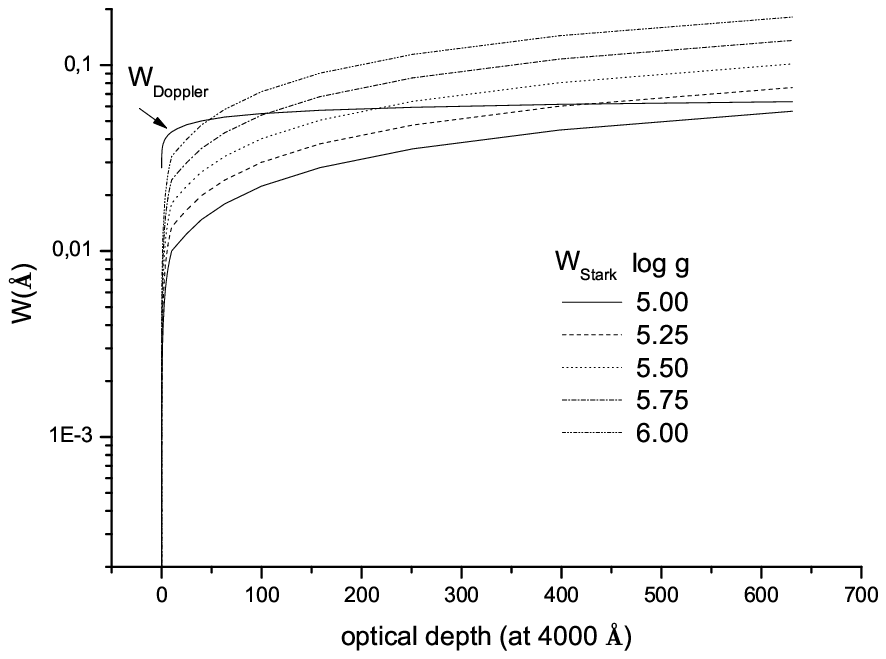}
\end{center}
\caption{Stark and Doppler widths for Ar III 4p $^{5}$P$_{2}$ - 5s
$^{5}$S$_{2}^{\rm{o}}$ ($\lambda$ = 2170.22 \AA) spectral line as
a function of optical depth. Stark widths are shown for five
values of model gravity log \emph{g} = 5-6, T$_{eff}$ = 22 000 K.}
\label{sdb2}
\end{figure}

Subdwarf B (sdB) stars are low-mass core helium burning stars with
extremely thin hydrogen envelopes located on the extreme
horizontal branch of the H-R diagram. The sdB stars have a high
effective temperature (20 000 K $\leq$ T$_{eff}$ $\leq$ 40 000 K)
and gravities (log $g$ $\simeq$ 5 - 6) (see e.g. \citet{Ohl00}).
Ar III spectral lines are observed in subdwarf B star atmospheres
\citep{Otoole06,Blanchette08}.

In hot star atmospheres, besides electron-impact broadening (Stark
broadening), the important broadening mechanism is the Doppler
(thermal) one as well as the broadening due to the turbulence and
stellar rotation. Other types of spectral line broadening, such as
van der Waals, resonance and natural broadening, are usually
negligible. For a Doppler-broadened spectral lines, the intensity
distribution is not Lorentzian as for electron-impact broadening
but Gaussian, and the full half-width of the spectral lines may be
determined by the equation (see e.g. \citet{Konjevic99})
\begin{eqnarray}
W_{D}[\rm\AA]=7.16\times 10^{-7}\lambda \lbrack
\rm\AA]\sqrt{\frac{T[K]}{M_{Ar}}}
\end{eqnarray}
where atomic weight of argon is $M_{Ar}$= 39.948 au.

The importance of Stark broadening mechanism for the Ar III 4p
$^{5}$P$_{2}$ - 5s $^{5}$S$_{2}^{\rm{o}}$ ($\lambda$ = 2170.22
\AA) spectral line in atmospheric conditions of sdB stars is
studied. We use the atmospheric models of \citet{Jeffery01}
(http://star.arm\\.ac.uk/~csj/models/Grid.html) which are
plane-parallel line-blanketed model atmospheres for hot stars in
local thermal, radiative and hydrostatic equilibrium. The
considered atmospheres have the following composition: 0.001
helium, 0.99741 hydrogen and 0.00047 carbon and nitrogen.

In Figs. \ref{sdb1} and \ref{sdb2}, we show Stark and Doppler
widths for Ar III 4p $^{5}$P$_{2}$ - 5s $^{5}$S$_{2}^{\rm{o}}$
($\lambda$ = 2170.22 \AA) spectral line as a function of
atmospheric layer temperature and as a function of the optical
depth (at 4000 \AA) respectively. Stark widths are shown for five
values of model gravity log $g$ = 5-6, T$_{eff}$ = 22 000 K. As we
can see in Fig. \ref{sdb1}, for the atmosphere with log $g$ = 6,
Stark broadening is the dominant broadening mechanism for the
atmospheric layers for which the temperature is higher than 43 000
K. For the atmosphere with log $g$ = 5.75, Stark width is equal to
Doppler width for the atmospheric layer with temperature
T$\thickapprox$ 50 000 K. For the atmospheres with log $g$ = 5.50
and log $g$ = 5.25, Stark width is higher than Doppler one only
for deep atmospheric layers. For the atmosphere with log $g$ = 5,
Stark width became comparable to Doppler one for the deeper layer
of the atmosphere (at T = 67638 K, W$_{Stark}$ = 0.0565 $\rm{\AA}$
and W$_{Doppler}$ = 0.0639 \AA). One should take into account,
however, that even when the Doppler width is larger than Stark
width, due to different behaviour of Gaussian and Lorentzian
distributions, Stark broadening may be important in line wings.

\section{Conclusions}
In this work we have determined Stark widths for 32 spectral lines
of Ar III, our results are in relatively good agreement with many
experimental results. The better agreement is found with
\citet{Bukvic08}. The largest disagreement is found with
\citet{Platisa75}. Comparison between theoretical and experimental
results allows to improve both theories and experiments. Our
results show that the use of the Cowan code \citep{Cowan81} for
the determination of oscillator strengths needed for SCP
calculation of Stark widths is very useful when no experimental
data exist. Our study of the Stark broadening in sdB stars shows
the importance of this mechanism especially for the atmospheres
with high values of log $g$.

\section*{Acknowledgment}
This work has been supported by the Tunisian research unit
05/UR/12-04. This work is also a part of the project 176002
"Influence of collisional processes on astrophysical plasma line
shapes" supported by the Ministry of Education, Science and
Technological Development of Serbia. A part of this work is
presented in the 9$^{th}$ SCSLSA.



\end{document}